\documentstyle[aps,multicol,epsfig]{revtex}
\begin{document}
\draft
\title{Vortex Fluctuations in High-$T_c$ Films: Flux Noise Spectrum and
  Complex Impedance}
\author{\"{O}rjan Festin and Peter Svedlindh}
\address{Department of Materials Science, Uppsala University, SE-751 21
Uppsala, Sweden}
\author {Beom Jun Kim and  Petter Minnhagen}
\address {Department of Theoretical Physics,
Ume{\aa} University, SE-901 87 Ume{\aa}, Sweden}
\author{Radoslav Chakalov and Zdravko Ivanov}
\address{Department of Physics, Chalmers University of
Technology, SE-412 96 G\"oteborg, Sweden}

\preprint{\today}
\maketitle
\begin{abstract}
The flux noise spectrum and complex impedance for a 500 {\AA} thick
YBCO film are measured and compared with predictions for two dimensional
vortex fluctuations. It is verified that the complex impedance and the
flux noise spectra are proportional to each other, that the logarithm of
the flux noise spectra for different temperatures has a common tangent
with slope $\approx -1$, and that the amplitude of the noise decreases as
$d^{-3}$, where $d$ is the height above the film at which the magnetic flux is
measured. A crossover from normal to anomalous vortex diffusion is indicated by
the measurements and is discussed in terms of a two-dimensional decoupling.
\end{abstract}


\begin{multicols}{2}
Spontaneously created two-dimensional (2D) vortices drive the phase
transition
of Kosterlitz-Thouless (KT) type between the
superconducting and the normal state for thin film
superconductors~\cite{minnhagen_rev}.
This means that the physics of the
vortices is responsible for the features in a region
around the transition~\cite{minnhagen_rev}.
More recently it has been found that high-$T_c$ superconductors also
show
behavior characteristic of 2D vortex fluctuations
~\cite{minnhagen_rev,norton,rogers,miu,persico}.
This is expected for films thin enough to be in the  "quasi" 2D
regime, while the phase transition for thicker films
are related to the 3D
bulk transition. However, also for such thick films 2D vortex
fluctuations have been
observed just above the transition~\cite{miu,persico}, which has been
ascribed to a
decoupling of the superconducting planes associated with the CuO$_2$
layers~\cite{minnols,friesen,pierson}.
The evidence in case of BSCCO, which has a fairly broad resistive
transition, is better established than for YBCO which has a very narrow
transition~\cite{miu,persico}. One may argue that for BSCCO, since it
has a
much larger anisotropy than YBCO, the interlayer coupling should be much
weaker and as a consequence the "quasi" 2D character much stronger
~\cite{minnhagen_rev}. From this perspective the possibility and
the significance of 2D vortex fluctuations remains to be clarified for
YBCO. In the present Letter we address this question by focusing on the
dynamical features of the vortices reflected in the flux noise spectrum and
the
complex impedance for a 500 {\AA} thick YBCO film;
a strategy which was also used in Ref.~\cite{rogers} for a "quasi"
2D BSCCO film. To resolve the narrow resistive region for our YBCO
sample, we use a very high temperature resolution in the experimental set-up.

The sample is a square shaped ($5 \times 5$ mm$^{2}$) 500 {\AA} thick
YBCO-123 film grown
epitaxially on a LaAlO$_{3}$ substrate by pulsed laser deposition.
According to X-ray $\theta$-2$\theta$ and $\phi$ scans the film is
highly
$c$-axis oriented and
only [100] YBCO $//$ [100] LaAlO$_{3}$  in-plane orientation is
observed. The experimental set-up is a SQUID based system designed for complex
impedance and flux noise measurements.
The sample space is magnetically shielded by $\mu$-metal and niobium
cans, resulting in a residual
dc field of approximately 1 mOe. The pick-up coil having
a diameter of 1.2 mm is a first order
gradiometer, each section containing 2 layers (8+7 turns) of 0.05 mm
NbTi wire. The amplitude of the field  in the complex impedance measurements
is 0.4 mOe at the center of the drive coil, which is low enough to ensure a
linear sample response. The construction of the system allows a very high
precision temperature control of the sample. The long time stability
of the sample temperature is of the order 0.1 mK \cite{festin}.

There exist predictions for the flux noise spectrum associated  with 2D
vortex fluctuations ~\cite{houlrik,hwang,tiesinga,fazio,timm}. In the
present analysis we focus on the relation between the flux noise spectrum
and the conductance: The flux noise spectrum $S(\omega)$ caused by 2D
vortex fluctuations should be proportional to the conductance $G(\omega)$
in the
limit where the distance $d$ between the sample and the pick-up coil is larger
than the perpendicular penetration depth $\Lambda$~\cite{kim}:
\begin{equation}
  S(\omega)=A{\rm Re}[G(\omega)]=\frac{A\rho_0(T)}{\omega} \left|{\rm
Im}\left[
  \frac{1}{\epsilon(\omega)}\right]\right|.
\label{S}
\end{equation}
Here $1/\epsilon(\omega)$, which is the quantity often studied
theoretically,
is the vortex dielectric function~\cite{minnhagen_rev}
and  $\rho_0$ is the bare superfluid density which within a
Ginzburg-Landau description decreases to zero as $\rho_0(T)\propto
1-T/T_{c0}$~\cite{minnhagen_rev}.
The proportionality constant $A$, apart from trivial
constants, contains a factor $T f(d/R)/R^2$,
where $R$ is the linear dimension of the pick-up coil.
The dissipation is proportional to $|{\rm
Im}\left[1/\epsilon(\omega)\right]|$ and
has a maximum as a function of $\omega$ at a fixed
$T$~\cite{minnhagen_rev}.
The frequency $\omega_0$ at this maximum defines the
characteristic
frequency scale which vanishes as the transition is
approached from above due to critical slowing down~\cite{hwang,jonsson}.

\begin{figure}
\centerline{\epsfxsize=8cm \epsfbox{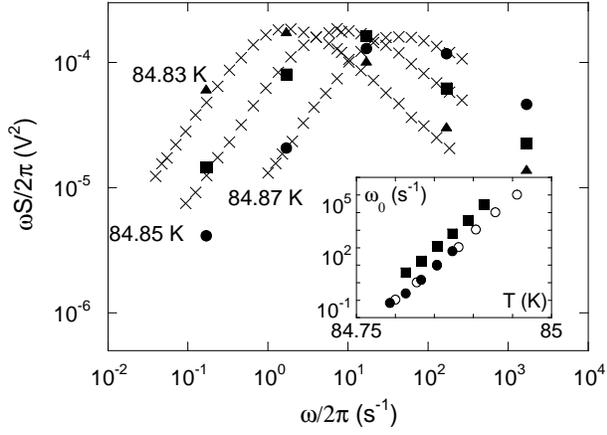}}
\vskip 0.5cm
\caption{Flux noise $\omega S(\omega)$
(crosses) and ac conductance ${\rm Re}[\omega G(\omega)]$
(filled circles, squares and triangles)
for three different $T$ plotted in a log-log scale.
The ac conductance results have been adjusted in the vertical direction.
The flux noise results are given in units of the SQUID voltage.
The inset shows the characteristic frequency $\omega_0$
obtained from the maxima of $\omega S(\omega)$ (filled circles) and from
the maxima in ${\rm Re}[\omega G(\omega)]$ vs. $T$ for
fixed $\omega$ [open circles, compare Fig. 3(b)]. Filled squares correspond to
the resistance $\cal{R}$ calculated as the inverse of the magnitude of
the small frequency plateaus of the noise spectra.
}
\label{fig1}
\end{figure}

In Fig.~\ref{fig1} we plot $\omega S(\omega)$ and ${\rm Re}[\omega
G(\omega)]$ for three temperatures above $T_c$:
They agree very well in all three cases which shows that
$S(\omega)\propto {\rm Re}[G(\omega)]$ as expected
for vortex fluctuations in the large $d$ limit.
The inset shows the corresponding characteristic frequency $\omega_0$
\{filled and open circles correspond to the maximum of $\omega S(\omega)$
and the dissipation peak of ${\rm Re}[\omega G(\omega)]$ vs. $T$
[compare Fig.~\ref{fig3}(b)], respectively\}. One notes the excellent
agreement between the characteristic frequencies for $S(\omega)$ and
$G(\omega)$ measured independently. A striking
feature in the inset is the dramatic decrease of $\omega_0$ [ $O(10^4)$
over a 0.1K interval] and a linear extrapolation gives $10^{-4}$ Hz at
84.7 K (which we take as a rough estimate of $T_c$). This rapid decrease of
$\omega_0$ is indeed expected from 2D vortex fluctuations because these imply
that $\omega_0$ is proportional
to the resistance $\cal{R}$~\cite{res}, which has a similar rapid
decrease, as is also seen in the inset of Fig.~\ref{fig1}.

The proportionality between the flux noise spectrum and the conductance
is by no means a general property. For example the flux noise spectra
discussed in Refs.~\cite{hwang,tiesinga} are
not proportional to the conductance~\cite{kim}, because they
correspond to the case when the distance $d$ from the sample to the pick-up
coil
is zero. Our experiment corresponds to $d$ larger than $\Lambda$ and only in
this limit is the  proportionality expected for 2D vortex
fluctuations~\cite{kim}.

\begin{figure}
\centerline{\epsfxsize=7cm \epsfbox{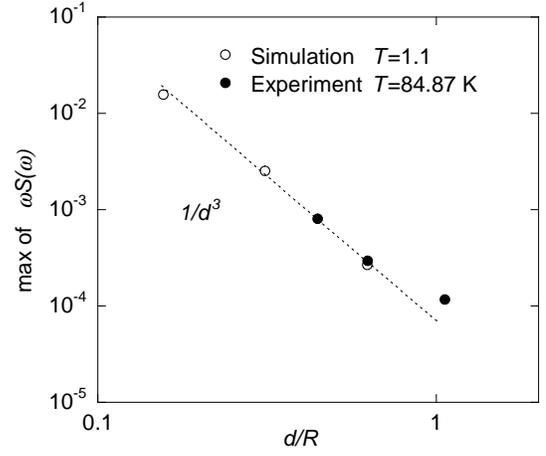}}
\vskip 0.5cm
\caption{Maximum of $\omega S(\omega )$ as a function of distance
$d$ to the pick-up coil in a log-log plot. The horizontal axis is $d/R$,
where $R$ is
the linear dimension of the pick-up coil. The open circles give results
from simulations (see Ref.~\protect\onlinecite{kim}) slightly above $T_c$.
The filled circles give the corresponding experimental results
(for convenience shifted in the vertical direction).}
\label{fig2}
\end{figure}

\begin{figure}
\centerline{\epsfxsize=8cm \epsfbox{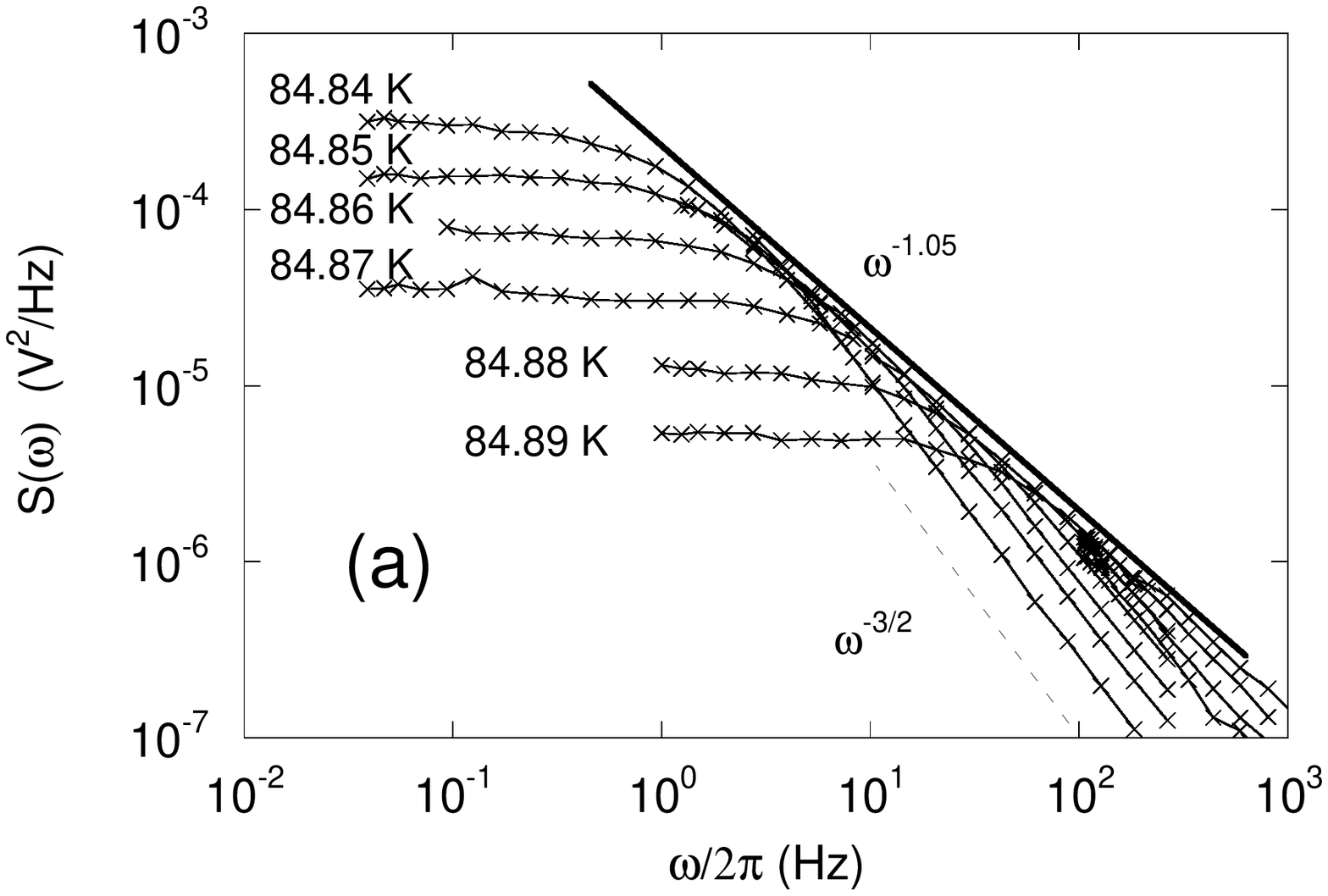}}
\centerline{\epsfxsize=8cm \epsfbox{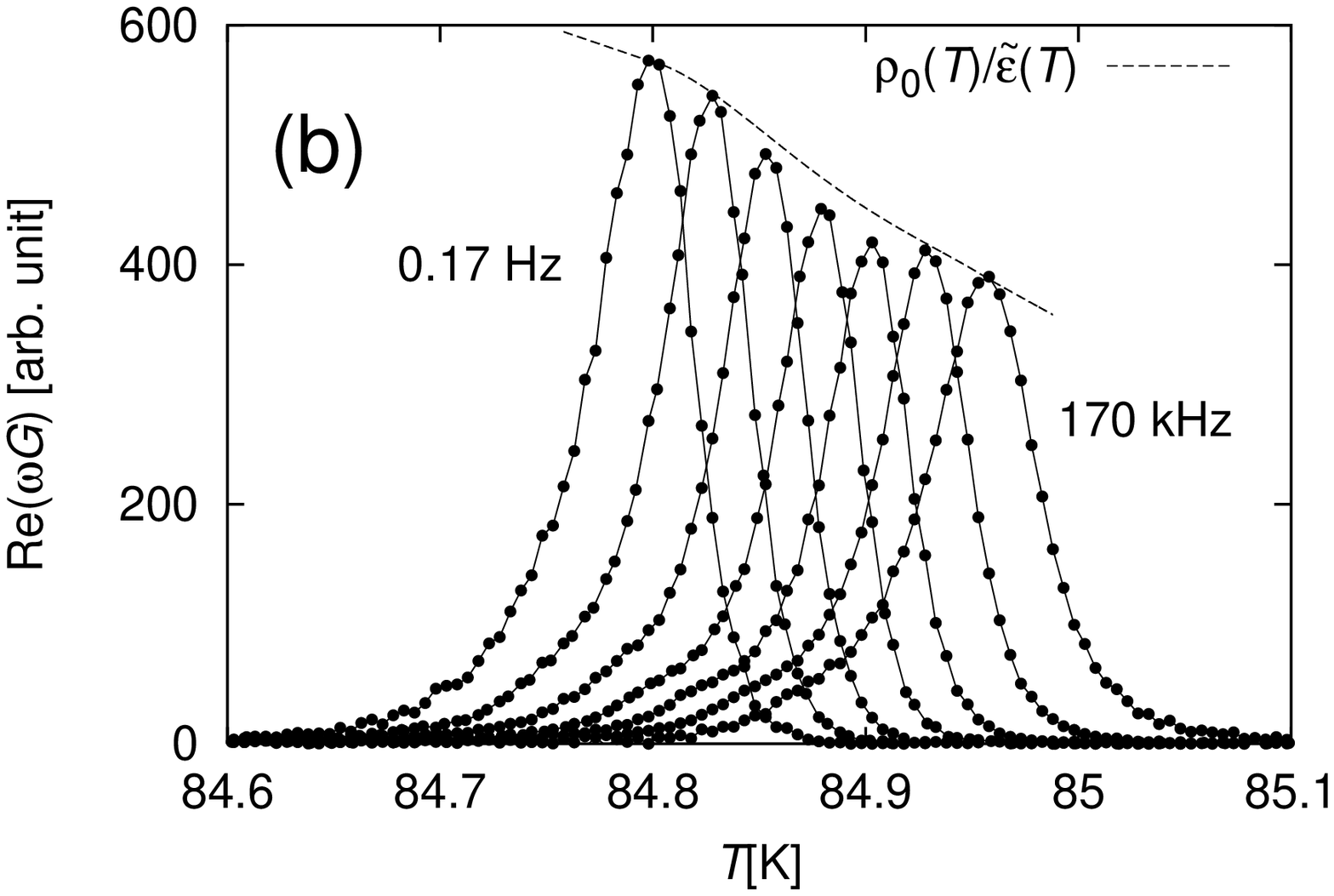}}
\caption{(a) Flux noise spectra for six temperatures
in a log-log plot has a common tangent (full drawn line)
with a slope $-1.05$. The slopes for the
individual spectra are close to $-3/2$ for larger frequencies
(dashed line). The flux noise is given in units of SQUID voltage.
(b) ${\rm Re}\left[\omega G(\omega)\right]$ as a function of $T$ for
different frequencies. The dashed curve corresponds to a
self-consistent determination of $\rho_0(T)/\tilde{\epsilon}(T)$
(see discussion of Fig.~4).
}
\label{fig3}
\end{figure}

A further test of the prediction for the flux noise
spectra when $d>\Lambda$ is the
dependence of the flux noise amplitude on $d$. For a fixed temperature
this dependence has the form $f(d/R)\propto 1/d^3$ for $R\gg d$ and
$1/d^6$ for $R\ll d$ [See Eq.~(\ref{S})]~\cite{kim}. Figure~\ref{fig2}
shows simulation results in the
region $R\approx d$ (open circles) and demonstrates that the $1/d^3$
behavior is valid also in this parameter range. We have obtained
experimental data in the same range (filled circles), which are in reasonable
agreement with the expected $1/d^3$ behavior~\cite{d}.

Since $\omega S(\omega)$ is proportional to
$|{\rm Im}\left[1/\epsilon(\omega)\right]|$,
properties of the vortex dielectric function
$1/\epsilon(\omega)$ should be reflected in the flux noise.
The connection between the vortex dynamics and $1/\epsilon(\omega)$
was worked out in Ref.~\cite{ambeg}. In the high-temperature phase
$1/\epsilon(\omega)$ for 2D vortex fluctuations close to $T_{c0}$ is of
normal Drude type describing the response of ``free''vortices:
\begin{equation}
 \left|{\rm
Im}\left[\frac{1}{\epsilon(\omega)}\right]\right|=\frac{1}{\tilde{\epsilon}}
\frac{\omega \omega_0}{\omega^2+\omega_0^2} ,
\end{equation}
where $\tilde{\epsilon}$ describes an effective static polarization of
vortex-antivortex pairs. The maximum value is $1/2\tilde{\epsilon}$ and
the peak ratio defined as ${\rm Im}[1/\epsilon(\omega)]/|{\rm Re}
[1/\epsilon(\omega)]$ at this maximum is 1. Close to $T_{c}$ the
response is expected to be dominated by bound vortex pairs and has been
found to be well parameterized by the Minnhagen
phenomenology (MP) form~\cite{minnhagen_rev,rogers,jonsson}:
\begin{equation}
\left|{\rm Im}\left[\frac{1}{\epsilon(\omega)}\right]\right|
=\frac{1}{\tilde{\epsilon}}\frac{2}{\pi}\frac{\omega \omega_0\ln
\omega/\omega_0}{\omega^2-\omega_0^2},
\end{equation}
which has the maximum value $1/\pi\tilde{\epsilon}$ and the peak ratio
$2/\pi$~\cite{jonsson,bmo}. In reality the response crosses
over from the anomalous MP form to the normal Drude form as the
temperature is increased from $T_c$ to $T_{c0}$~\cite{jonsson,bmo}.
This is because at $T_{c0}$ there are only "free" vortices and no bound
vortex-antivortex pairs as a consequence of the thermal unbinding of
vortex pairs. However, as the temperature is lowered the proportion of bound
pairs increases and for the pure 2D case these bound pairs dominate the
response in a region above the 2D transition~\cite{minnhagen_rev}. This
knowledge
of $1/\epsilon(\omega)$ together with the proportionality
to $\omega S(\omega)$ leads to the prediction of a common tangent for
the flux noise spectra for different temperatures~\cite{jonsson}:
\begin{equation}
\omega S(\omega )|_{\omega_0} \propto
\frac{T\rho_0(T)h(T)}{\tilde{\epsilon}(T)} .
\end{equation}
If the right hand side is independent of $T$ then
$S(\omega_0)$ for different $T$ lie on a common curve $\propto 1/\omega$
such that all spectra $\ln S(\omega)$ as a function of $\ln \omega $ has
a common tangent
with slope $-1$. The point is that, although the right hand side does
have a weak $T$-dependence ($\rho_0\propto 1-T/T_{c0}$, the height
$h(T)$ of $\tilde\epsilon {\rm Im}(1/\epsilon)$ goes
from $1/\pi$ to $1/2$ as $T$ goes from $T_c$ to $T_{c0}$, and
$\tilde{\epsilon}$
is of order unity and goes from a value $>1$ at $T_{c}$ to 1 at
$T_{c0}$~\cite{minnhagen_rev}),
the dramatic $T$-dependence of $\omega_0(T)$ dominates,
leading to  the common tangent with the slope $\approx -1$.
It should be noted that the common tangent only hinges on two things,
the proportionality between
$\omega S(\omega)$ and $|{\rm Im}(1/\epsilon(\omega))|$ in the large
$d$ limit and the weak $T$-dependence of the proportionality factor in
comparison
with the large temperature dependence of the characteristic frequency
$\omega_0$, but it does not hinge on the explicit
frequency dependence of $|{\rm Im}(1/\epsilon(\omega))|$.  One may note
that the flux noise  spectra for $d=0$ (discussed, e.g., in
Refs.~\cite{hwang,tiesinga})
are not proportional to the conductance and, as a consequence, do not
have a common tangent. Figure~\ref{fig3}(a) tests this prediction for the flux
noise spectra for six different $T$:
A common tangent is indeed found and has the slope $-1.05$ very close to
$-1$. As also seen in Figure~\ref{fig3}(a) each individual noise spectrum has
a slope  $-3/2$ for frequencies $\omega>\omega_{0}$.
Because of the proportionality shown in Fig.~\ref{fig1}, this also means
a corresponding slope -3/2 for ${\rm Re}[G(\omega)]$. But ${\rm
Re}[G(\omega)]$ for a superconductor must eventually have a crossover to an
$\omega^{-2}$-dependence for high enough frequencies. However, this crossover
frequency is outside our present experimental range. By comparison
the simulations in Ref.~\cite{kim} for the flux noise spectrum from 2D
vortex fluctuations and correspondingly the conductance show a slope close -3/2
only over a limited frequency region before the crossover to -2 sets in.

\begin{figure}
\centerline{\epsfxsize=10cm \epsfbox{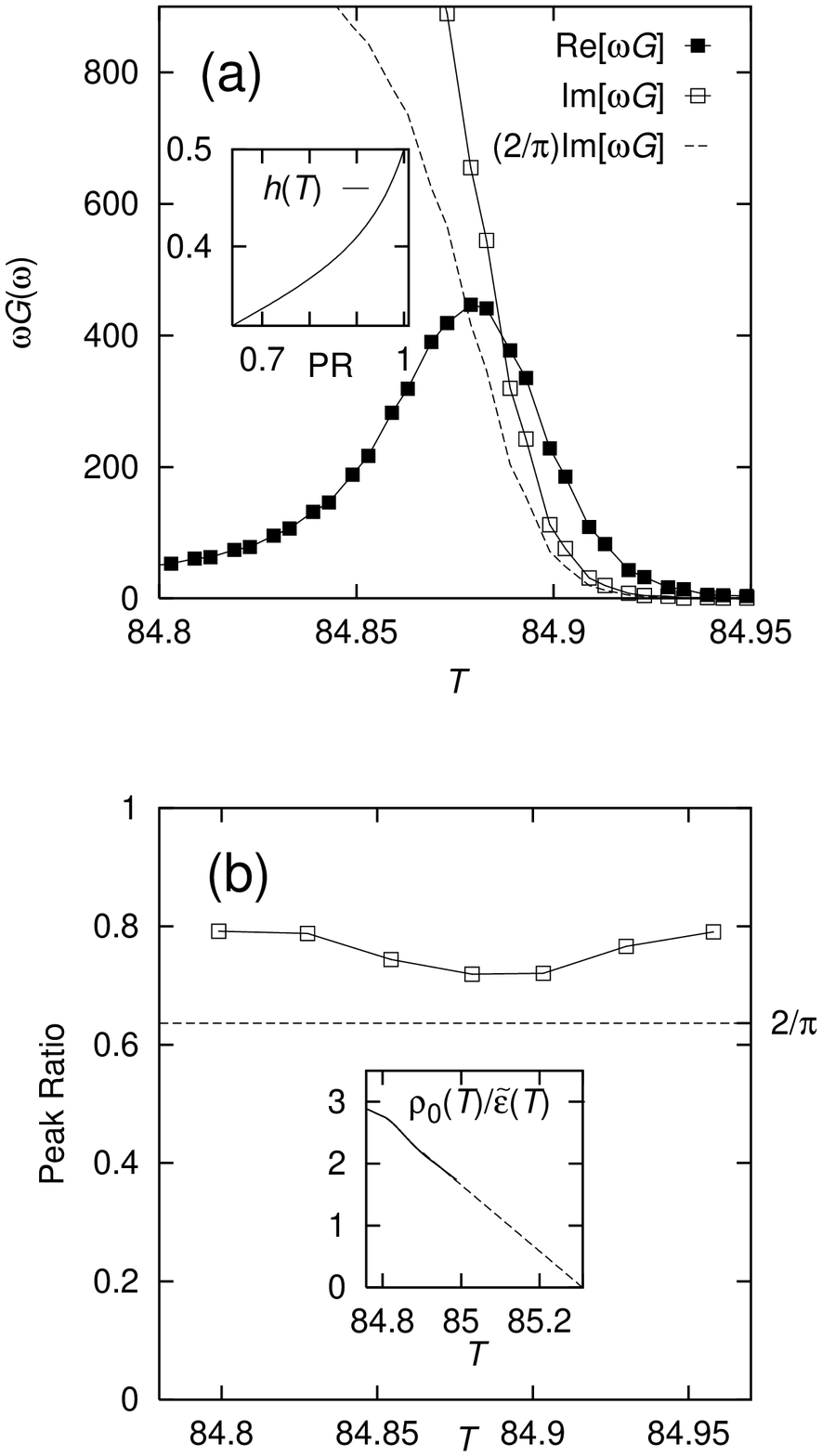}}
\caption{(a) $\omega G(\omega)$ as a function of
$T$ for the frequency $\omega/2\pi=170$ Hz (filled squares for real part
and
open squares for imaginary part). The dashed curve is $C \times {\rm
Im}[\omega
G(\omega)]$, where $C\approx \pi/2$ is the peak ratio (see text for
explanation).
The inset shows the relation between the peak height $h(T)$ and the peak
ratio (PR).
(b) The peak ratios as a function of $T$ obtained from the
self-consistent determination described in text. The inset shows that
the
self-consistently
determined $\rho_0(T)/\tilde{\epsilon}(T)$ is proportional to
$1-T/T_{c0}$ which gives $T_{c0}\approx 85.3$ K.}
\label{fig4}
\end{figure}

We can also use the knowledge of $1/\epsilon(\omega)$ when analyzing
the complex conductance $G(\omega)$.
Figure~\ref{fig3}(b) shows ${\rm Re}[\omega G(\omega)]$, which is
proportional to $|{\rm Im}[1/\epsilon(\omega)]|$, as a function of $T$
for several frequencies.
If the proportionality factor $\rho_0(T)/\tilde{\epsilon}(T)$ was
independent of $T$, then the frequency at the
maximum of ${\rm Re}[\omega G(\omega)]$ in Fig.~\ref{fig3}(b) for each
$T$ would give the corresponding
characteristic frequency $\omega_0(T)$ of $|{\rm
Im}[1/\epsilon(\omega)]|$. Since the $T$-dependence of
$\omega_0(T)$ completely dominates over the $T$-dependence of the
proportionality factor this gives a very good estimate
(this is the estimate used in the inset of Fig.~\ref{fig1}). In case of
$G(\omega)$ we can push the analysis one step further by using the
relation between the peak ratio and the height $h(T)$ for
$\tilde\epsilon{\rm Im}[1/\epsilon(\omega)]$ inferred from simulations
of 2D vortex fluctuations~\cite{bmo}.
This relation is shown in the inset of Fig.~\ref{fig4}(a), as obtained
from the parameterization given in Ref.~\onlinecite{bmo}.
Figure~\ref{fig4}(a) shows ${\rm Re}[\omega G(\omega)]$
and ${\rm Im}[\omega G(\omega )]$ for $\omega/2\pi=170$ Hz.
The dashed line is $C\times {\rm Im}[\omega G(\omega)]$,
where the constant $C$ is determined so that the dashed curve cuts
${\rm Re}[\omega G(\omega)]$ precisely at
the peak. This gives $C\approx 2/\pi$ consistent with bound
vortex-antivortex pair response described by Eq. (3) and consequently with
$h(T)=1/\pi$ for the $T$ corresponding to the maximum. This construction
can be further improved by taking the $T$-dependence of
$\rho_0(T)/\tilde{\epsilon}(T)$ into account:
The quantity corresponding to $1/\epsilon(\omega)$ is $\tilde{\epsilon}\omega
 G(\omega) /\rho_0$.
We start from the $T$-dependence for $\rho_0/\tilde{\epsilon}$
corresponding to a $T$-independent $h(T)$. From this one gets the peak
ratio for
each frequency and hence $h(T)$ for each $T$ corresponding to a maximum.
Using this obtained $h(T)$ one gets a new estimate for
$\rho_0/\tilde{\epsilon}$.
In this way we can self-consistently determine the $T$-dependence of
$\rho_0/\tilde{\epsilon}$ and the peak ratio [both shown in
Fig.~\ref{fig4}(b)].
The inset in Fig.~\ref{fig4}(b) shows that $\rho_0(T)/\tilde{\epsilon}$
goes linearly to zero
at $T_{c0}\approx 85.3$ K. This is consistent with the Ginzburg-Landau
prediction
$\rho_0(T)\propto 1-T/T_{c0}$ and the 2D Coulomb gas prediction
$\tilde{\epsilon}\approx 1$ for $T$ slightly above
$T_c$~\cite{minnhagen_rev}.

The peak ratios in Fig.~\ref{fig4}(b) suggest an increase towards the
Drude
value 1 as $T_{c0}$ is approached and a decrease towards the bound
vortex-antivortex value $2/\pi$
as the temperature is decreased. This is consistent with the expected
crossover from normal to anomalous vortex diffusion for 2D vortex
fluctuations~\cite{minnhagen_rev,jonsson}. However, as the temperature
is
further decreased towards $T_c$ the peak ratio again increases.
This might be caused by a plane decoupling~\cite{minnhagen_rev,minnols}:
Closely above $T_c$ the vortex fluctuations induce a decoupling of the
YBCO
material into essentially decoupled superconducting CuO$_2$ layers.
The 2D vortex fluctuations associated with these planes are
responsible for the peak ratios close to the bound vortex-antivortex
value as well as the increase at higher
$T$. The coupling of the planes very close to $T_c$ suppresses the
vortex-antivortex pair fluctuations and
this we suggest could be the cause of the increase of the peak ratio
with decreasing $T$ since the peak ratio depends on the relative
proportion of free and bound vortices~\cite{jonsson1}.

In summary we have measured the flux noise and the complex
impedance for a YBCO film and shown that in a narrow temperature
interval just above the resistive transition the dynamical features
can be consistently interpreted in terms of 2D vortex fluctuations.
To this end we explicitly verified the proportionality between
flux noise and the complex conductance valid when the distance $d$ to
the pick-up coil is large, as well as the $1/d^3$ dependence
of the flux noise amplitude. The connection to the vortex dielectric
function gives rise to a common tangent with a slope
$\approx -1$ for the flux noise at different temperatures which was
also verified.
Finally, we self-consistently determined the peak ratio associated
with the vortex dielectric function and suggested an interpretation
in terms of a plane decoupling just above the resistive transition. This
work suggests that 2D vortex fluctuations play an identifiable role even
in an YBCO thin film sample with a very narrow zero field resistive
transition.

The research was supported by the Swedish Natural Science Research
Council
through Contract Nos. FU 04040-322 and FU 04657-347.

\narrowtext

\vskip 10cm

\newpage

\end{multicols}


\begin{thebibliography}{99}
\bibitem{res} We estimate $\cal{R}$ as $\cal{R}$ $=1/G(0)$ and note that
$G(0)$
corresponds to the magnitude of the small frequency plateau of the noise
spectrum.
\bibitem{d} In the experimental set-up the section of the pick-up coil
being closest to the film has a vertical extent of 0.45 mm. $d$ is
determined as the weighted average over this coil, $d=\Sigma_i d_i\times
p(d_i)$ with
$p(d_i)=(1/d_i^3)/\Sigma_j (1/d_j^3)$.
\end{thebibliography}
\end{document}